\documentclass[letterpaper]{article} 
\usepackage{aaai2027}  
\usepackage[hyphens]{url}  
\usepackage{graphicx} 
\usepackage{natbib}  
\usepackage{caption} 
\usepackage{algorithm}
\usepackage{algorithmic}

\usepackage{newfloat}
\usepackage{listings}
\DeclareCaptionStyle{ruled}{labelfont=normalfont,labelsep=colon,strut=off} 
\floatstyle{ruled}
\newfloat{listing}{tb}{lst}{}
\floatname{listing}{Listing}

\usepackage{booktabs}
\usepackage{amsmath}
\usepackage{amssymb}
\usepackage{multirow}

\title{From Generation to Discovery: Diffusion Mutation Kernels for Circuit and Physical Design}
\author{
Dinithi Jayasuriya,\textsuperscript{\rm 1}
Aravind Saravanan,\textsuperscript{\rm 1}
Nilesh Ahuja,\textsuperscript{\rm 2}
Amanda Rios,\textsuperscript{\rm 2}
Amit Trivedi\textsuperscript{\rm 1}
}

\affiliations{
\textsuperscript{\rm 1}University of Illinois Chicago, Chicago, IL, USA\\
\textsuperscript{\rm 2}Intel Corporation, USA
}

\begin{document}

\maketitle

\begin{abstract}
Generation and discovery are different problems. A generative model trained on valid artifacts reproduces a distribution, whereas discovery must produce artifacts that lie outside the observed corpus, satisfy hard structural constraints, and improve on established designs under evaluation that the model cannot influence. We introduce SteerGenSE, a diffusion-based discovery framework. Unlike conventional generative models that sample from learned distributions, SteerGenSE learns transition operators that transform existing artifacts into new candidates. Controlled partial re-noising followed by denoising defines a diffusion mutation kernel, a learned transition distribution that preserves the structural regularities of feasible designs while moving between regions of the design space. The learned model supplies feasibility structure only, and all correctness and performance judgments remain with external engineering tools. Intermediate diffusion trajectories are additionally monitored under a conformal risk budget so that unpromising candidates are discarded before expensive evaluation. We evaluate the framework on three electronic design spaces, an environment that supplies rigorous non-differentiable evaluators in the form of simulation, formal equivalence checking, and industrial physical implementation. The framework discovers 32-bit prefix adders that are formally verified equivalent to addition over all $2^{64}$ input pairs and reduce delay by $17\%$ and area by $18\%$ relative to Kogge-Stone under a placed-and-timed flow; seven independently re-simulated amplifier topologies absent from the training corpus, spanning gains of $21.9$-$66.1$\,dB and bandwidths of $72.9$\,kHz-$207$\,MHz; and macro placements on held-out netlists reaching $0.68\times$ wirelength of an industrial placer.
\end{abstract}

\begin{figure*}[t]
\centering
\includegraphics[width=0.85\textwidth]{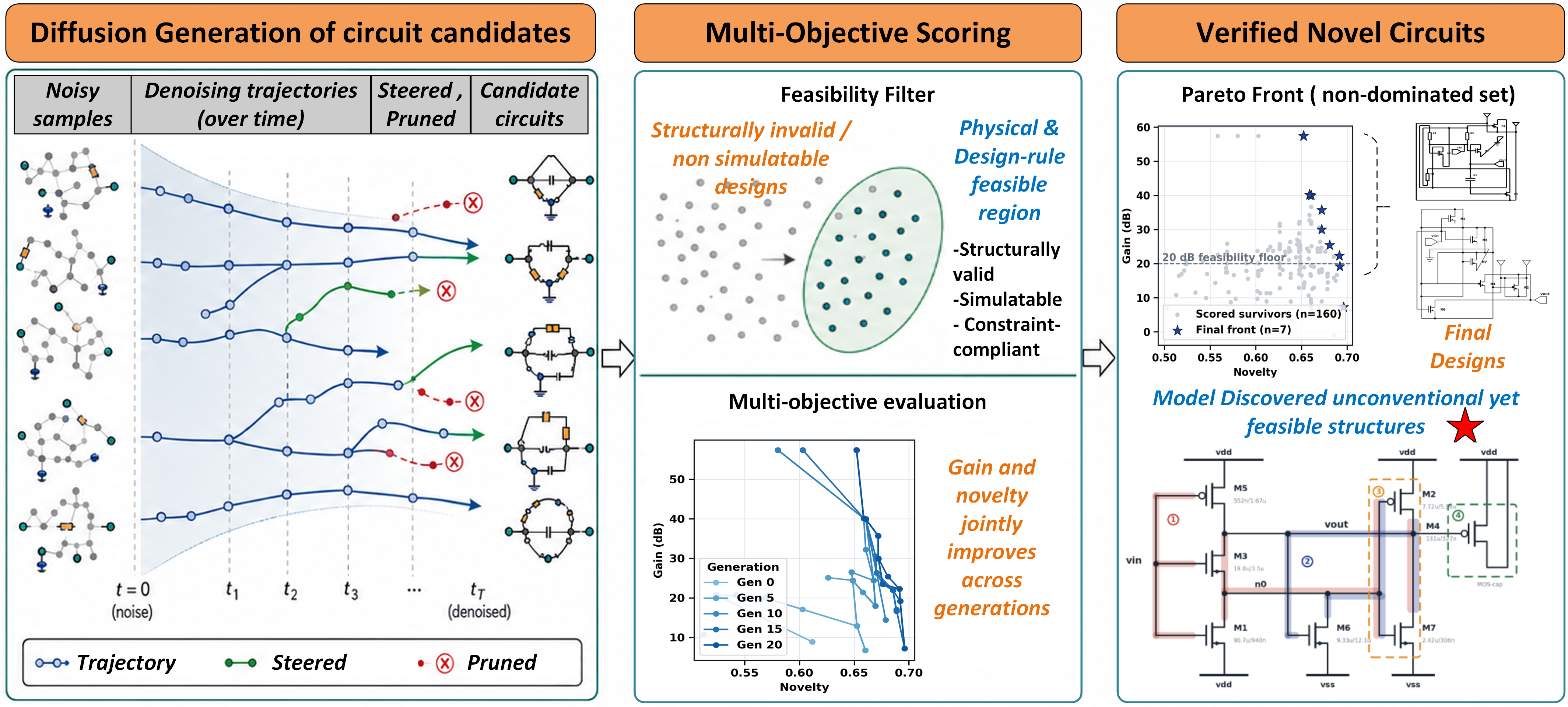}
\caption{Overview of SteerGenSE. A learned diffusion prior generates and mutates candidate designs, while trajectory-aware steering prunes unpromising paths and reallocates computation. External engineering tools evaluate feasibility, performance, and novelty, and Pareto selection determines the population carried into the next generation.}

\label{fig:framework-overview}
\end{figure*}

\section{Introduction}
\label{sec:intro}

Diffusion models have become a dominant approach for generating structured
artifacts, including images, molecules, proteins, and programs
\citep{ho2020denoising,vignac2022digress,hoogeboom2022equivariant}. Generation and
discovery, however, pursue different goals. Generation imitates a training
distribution through independent samples, whereas discovery iteratively searches
for artifacts beyond observed examples that satisfy hard constraints and deliver
utility under an evaluator the model does not control.

Three properties make design discovery challenging. First, valid design
manifolds are sparse: arbitrary circuit graphs do not simulate, prefix networks
do not compute addition, and macro placements violate physical constraints.
Second, novelty is harder than optimization. Retuning a Kogge--Stone variant
differs fundamentally from discovering a new prefix organization, just as
resizing a known amplifier differs from discovering an unseen topology. Third,
evaluation is expensive and black box. Candidates cannot be enumerated, no
gradient passes through SPICE, OpenROAD, or SAT solvers
\citep{wolf2013yosys}, and learned reward models are unreliable substitutes.
A discovery system must therefore preserve feasibility, seek novelty, and obtain
external evidence of utility simultaneously.

Existing paradigms address these requirements separately. Generative models
learn where valid artifacts exist but do not incorporate evaluation feedback
into subsequent proposals. Evolutionary and surrogate methods iterate through
handcrafted variation operators or learned objective approximations that can be
exploited rather than improving true design quality
\citep{holland1992adaptation,trabucco2021conservative}. This raises the question:
Can a generative model learn both where valid designs exist and how they evolve?

Our answer builds on the observation that reverse diffusion is not only a
decoder from noise, but also a learned transition mechanism between valid
artifacts. Re-noising a design to a chosen depth and denoising it again produces
a candidate that retains parent structure while exploring directions learned
from data. We call this conditional distribution a diffusion mutation kernel.
It converts a trained diffusion model into a search operator over its support
without additional training. Figure~\ref{fig:framework-overview} illustrates
the framework, where external tools provide all feasibility and performance
judgments, while intermediate trajectories are monitored under a conformal risk
budget to discard unpromising candidates.

We use circuit design as a benchmark environment since it provides discrete and continuous search spaces, hard validity
constraints, and mature evaluators with non-differentiable, model-independent
feedback. Across three spaces, SteerGenSE discovers 32-bit prefix adders that
are formally verified before and after technology mapping and improve on
Kogge-Stone by $17\%$ in delay and $18\%$ in area under a placed-and-timed flow,
seven re-simulated amplifier topologies absent from the training corpus, and
macro placements on held-out netlists reaching $0.68\times$ the wirelength of
an industrial placer. Our contributions are:

\begin{itemize}

\item \textbf{Diffusion mutation kernels} that turn diffusion models from
artifact generators into learned search operators.
\item \textbf{External-evaluator-driven discovery}, where the model proposes
structure and engineering reality judges quality.
\item \textbf{A cross-domain discovery framework}, instantiated without change
across discrete topology, symbolic architecture, and continuous spatial design
spaces.
\item \textbf{Risk-controlled steering}, which cuts black-box
evaluations under a conformal bound on the risk of discarding a successful
trajectory, and the conditions where it helps.
\end{itemize}

\section{Relation to Prior Work}

\vspace{2pt}
\noindent\textbf{Generative models for design.}
Generative design approaches learn distributions over valid artifacts and
sample from them
\citep{gao2025analoggenie,chang2024lamagic,dong2023cktgnn,li2025analogfed,kim2026analogtobi}.
They learn $p(x)$ and draw $x$. We learn $p(x)$ and use the same network to
realize $p(x'|x)$, a distribution over how one valid artifact becomes another.
Only the second supports search in which evaluation outcomes shape later
proposals.

\vspace{2pt}
\noindent\textbf{Evolutionary search.}
Population-based methods explore through selection and variation
\citep{deb2002fast,hansen2001completely}, but their mutation operators encode
manually specified assumptions about how artifacts may change, tied to a
particular encoding. We retain population-based selection but learn the
transition distribution from valid artifacts for transferability across spaces.

\vspace{2pt}
\noindent\textbf{Surrogate optimization.}
Bayesian optimization, offline model-based optimization, GFlowNets, and
reinforcement learning for design
\citep{jones1998efficient,trabucco2021conservative,bengio2021flow,mirhoseini2021graph,roy2021prefixrl}
improve efficiency by approximating objectives or rewards, which biases
exploration toward regions where the approximation is inaccurate. Our learned
model never approximates the objective. It represents the structural
regularities of feasible designs and delegates every performance decision to an
external evaluator, so artifacts are validated only by executable engineering
flows.

\vspace{2pt}
\noindent\textbf{Diffusion mechanisms and evaluation environments.}
We build on discrete diffusion, classifier guidance, and partial denoising
\citep{vignac2022digress,austin2021structured,ho2022classifier,meng2021sdedit} and
use them as components of an iterative discovery process. Simulation,
implementation, and verification flows
\citep{ren2020paragraph,lee2024chip,trung2025diffplace,jung2021metrics2,lai2025analogcoder}
validate discovered artifacts beyond distributional similarity, and we use them
as external judges of a discovery method.

\section{Diffusion as a Learned Search Operator}
\label{sec:framework}

\noindent\textbf{Discovery as search over learned design spaces.}
Let $x$ denote a design artifact and $c$ a condition specifying the desired requirement. We formulate discovery as multi-objective optimization $F(x)=\left(f_1(x),f_2(x)\right)$, where $f_1(x)$ is performance measured by an external engineering tool and $f_2(x)$ is novelty relative to existing designs. The objective is $\max_x F(x)\ \text{s.t.}\ g(x)=1$, where $g(x)$ is a hard validity predicate ranging from structural constraints to formal verification. Because discovery balances novelty against utility, the solution is a Pareto set rather than a single optimum.

Effective discovery requires a learned representation of where valid artifacts exist and a mechanism for exploring transformations within that region without manually specified rules. A diffusion model provides both.

\vspace{2pt}\noindent\textbf{Learning the feasible design space.}
We learn the structural regularities of valid designs rather than encoding domain-specific rules. A conditional diffusion model is trained on valid artifacts $\mathcal{D}$ with distribution $p_{\theta}(x_0|c)$. The learned design space is $\mathcal{M}_{\theta}(c)=\operatorname{supp}p_{\theta}(\cdot|c)$, a prior over regions where feasible artifacts are likely, and discovery becomes $\max_{x\in\mathcal{M}_{\theta}(c)}F(x)$ subject to $g(x)=1$.

The prior captures recurring structural organizations of realizable artifacts without handcrafted constraints or templates, and can produce configurations that respect them without appearing in the training corpus.

\vspace{2pt}\noindent\textbf{The diffusion mutation kernel.}
Discovery requires controlled transitions between artifacts, in which a new candidate preserves the structural dependencies that make a design realizable while exploring a different configuration. With $T$ denoising steps, we obtain such transitions by perturbing a parent design to step $\gamma T$ and applying the reverse process,
\begin{equation}
K_{\gamma}(x'|x)
=
\int q(x_{\gamma T}|x)
\prod_{t=\gamma T}^{1}
p_{\theta}(x_{t-1}|x_t,c)
dx_{1:\gamma T}.
\end{equation}

We call $K_{\gamma}$ the diffusion mutation kernel. Standard diffusion sampling produces independent draws from $p_{\theta}(x|c)$, whereas $K_{\gamma}$ defines a conditional transition distribution anchored at an existing artifact. This is where generation and discovery separate. The model learns not only where valid artifacts exist, but how valid artifacts evolve, and the second capability converts a generative model into a search operator without any additional training objective.

The two capabilities are complementary, which is why random exploration fails here. The feasible manifold occupies a small fraction of the ambient space, so unguided perturbation rarely lands on a realizable design. The prior confines exploration to structurally meaningful regions, and $K_{\gamma}$ decides which transitions inside them preserve realizability.

This changes what is assumed and what is learned. Evolutionary methods take the variation operator as given, so a human fixes in advance which moves are admissible. $K_{\gamma}$ makes the geometry of design evolution itself an object of learning, estimated from the distribution of valid artifacts. One operator consequently serves a motif graph, a categorical grid, and continuous coordinates without redesign.

Surrogate methods replace the evaluator with a learned approximation of the objective. The diffusion model never predicts objective values. It models feasible transitions only, and performance judgments stay with the external evaluator.

The parameter $\gamma$ controls the transition scale,
\begin{equation}
K_0(x'|x)=\delta(x'-x),
\qquad
K_1(x'|x)\approx p_{\theta}(x'|c),
\end{equation}
where small $\gamma$ favors local refinement and large $\gamma$ broader exploration. For the two discrete instantiations, the integral is a sum over states and $\delta$ is a Kronecker delta. Following Table~\ref{tab:gamma-sweep}, we use $\gamma=0.3$ for placement and prefix adders and $\gamma=0.5$ for analog circuits.

\vspace{2pt}\noindent\textbf{External-evaluation discovery loop.}
Our loop separates roles that are coupled in existing optimization frameworks. The diffusion model defines where and how to explore. Feasibility, correctness, and performance are established exclusively by external engineering flows for simulation, synthesis, implementation, and verification. Given a population $P_k$, offspring drawn from the mutation kernel are evaluated alongside their parents, and Pareto selection $S_{\mu}$ to population size $\mu$ produces the next generation,
\begin{equation}
P_{k+1}
=
S_{\mu}
\left(
P_k
\cup
\{x'\sim K_{\gamma}(\cdot|x):x\in P_k\}
\right).
\label{eq:master}
\end{equation}
This separation enables surrogate-free discovery.

\subsection{Trajectory-Aware Risk-Controlled Steering}
\label{sec:method-steering}

Equation~\eqref{eq:master} evaluates candidates only after denoising completes, which wastes computation on trajectories that become unpromising early. Steering complements the discovery loop by reducing that waste, while external evaluation remains the final authority on quality. Whether it pays for itself depends on the ratio between evaluator and generation cost and on whether intermediate trajectories give reliable failure signals. Both properties are measurable before search begins, and Section~\ref{sec:steering-when} analyzes them across domains.

We divide mid-generation control into two tasks. \emph{Pruning} stops unpromising trajectories, and \emph{reallocation} redirects the released computation. Only a conformally calibrated signal may prune, whereas an uncalibrated health heuristic is restricted to reallocation among survivors.

\vspace{2pt}\noindent\textbf{Two views of a trajectory.}
At checkpoint $t_k$ a trajectory exposes two signals, its current noisy state
$z^{(i)}_{t_k}$ and its predicted clean endpoint $\widehat{x}^{(i)}_{0,t_k}$,
and decoding both gives graphs $G^{\mathrm{cur}}_{i,k}$ and
$G^{\mathrm{pred}}_{i,k}$. They answer different questions. The current state
answers where a candidate is now. Because noisy states routinely recover during
later denoising, its health
\begin{equation}
h^{(i)}_k
=
\exp\!\left[
-w_c v_c\!\left(G^{\mathrm{cur}}_{i,k}\right)
-w_d v_d\!\left(G^{\mathrm{cur}}_{i,k}\right)
\right]
\label{eq:current-health}
\end{equation}
may only rank survivors for compute allocation and may never terminate them,
where $v_c$ counts hard structural violations and $v_d$ counts dangling
connections. The predicted endpoint answers where a candidate is heading, and
it is the only signal permitted to prune. Its forecast damage
$d_{i,k}=\operatorname{Dangling}(G^{\mathrm{pred}}_{i,k})$ estimates the defect
the trajectory is converging toward.

\vspace{2pt}\noindent\textbf{Conformal pruning.}
We prune on the predicted endpoint and calibrate that decision with conformal
risk control\citep{vovk2005algorithmic,angelopoulos2021gentle}. Forecast damage is normalized against trajectories that the
unpruned policy carried to a valid design, giving two running-maximum
statistics that capture different failure modes. Each is thresholded
independently at risk $\alpha/2$, and a trajectory is pruned at the first
checkpoint where either threshold is crossed (more details in
Appendix~\ref{app:calibration}).

\vspace{2pt}\noindent\textbf{Winner-preservation guarantee.}
Because both statistics are running maxima, any threshold crossing at
checkpoint $k$ also holds at completion. Under exchangeability between
calibration winners and a new winner generated by the same fixed policy,
conformal calibration bounds each test's winner-pruning probability by
$\alpha/2$, and the union bound gives
\begin{equation}
    \Pr\!\left(
        \text{prune}\mid Y_{\mathrm{new}}=1
    \right)
    \leq \alpha .
    \label{eq:winner-preservation}
\end{equation}
We use $\alpha=0.1$. The guarantee is policy-specific, so changes to the
generation, evaluation, checkpoint, or winner-definition protocol require
recalibration.

\vspace{2pt}\noindent\textbf{Compute reallocation.}
Released computation is reused by sequential Monte Carlo\citep{smith2013sequential},
which denoises $N$ particles in parallel and refills pruned slots by resampling
survivors on health $h^{(i)}_k$, capping clones per source to preserve diversity.
Under a matched budget of 128 SPICE evaluations this gives the best structural
validity and simulation success, so we adopt it as the analog generation backend.
Pruning without reallocation raises hypervolume but leaves validity and efficiency
near baseline, having freed computation without deciding where to reassign it.
Appendix~\ref{app:steering-ablation} (Table~\ref{tab:steering-ablation}) gives the
full sampler comparison, including tree search.

\section{Discovery Across Three Design Spaces}
\label{sec:instantiations}

We instantiate SteerGenSE in parallel-prefix adder synthesis, analog circuit topology discovery, and chip macro placement, evaluated using synthesis with formal verification, SPICE, and OpenROAD, respectively. Table~\ref{tab:domains} summarizes the three settings. The artifact representation, validity predicate, and evaluator differ in every case, while the discovery formulation and search procedure are unchanged.

\begin{table}[t]
\centering
\small
\setlength{\tabcolsep}{2.4pt}
\renewcommand{\arraystretch}{1.05}

\begin{tabular}{@{}lccc@{}}
\toprule
& Prefix adder
& Analog
& Placement \\
\midrule
Artifact
& categorical grid
& motif graph
& coordinates $\mathbb{R}^{n\times 2}$ \\

Diffusion
& discrete
& discrete
& continuous \\

Validity $g$
& SAT equivalence
& structural gate
& legalization \\

Evaluator $F$
& mapped timing
& \texttt{ngspice}
& OpenROAD \\

Discovery
& new structure
& new structure
& new organization \\
\bottomrule
\end{tabular}

\caption{One discovery framework across heterogeneous design spaces. The
artifact representation, validity mechanism, and engineering evaluator vary
across domains, while the discovery formulation and search process remain
unchanged.}
\label{tab:domains}
\end{table}
\subsection{Discovery of Novel Prefix Architectures}
\label{sec:prefix}

Our primary instantiation targets 32-bit parallel-prefix adders, a space in which functional correctness can be verified exhaustively. Each input bit generates a propagate--generate pair, and prefix operators hierarchically combine intervals to compute carries. The framework discovers the topology of the prefix network rather than operand values, so this domain tests whether generative discovery produces new functional architectures with formal correctness guarantees.

\vspace{2pt}\noindent\textbf{Representation of functional architectures.}
A prefix topology is a fixed-size categorical grid over levels and bit positions, where each occupied cell names the earlier bit position whose prefix is combined at that level. The grid decodes deterministically into a directed acyclic prefix graph, so search operates directly on architectural dependencies rather than on a surface encoding of them. The training corpus contains $16{,}269$ unique 32-bit structures, including ripple, Sklansky \citep{sklansky1960conditional}, Kogge-Stone\citep{kogge1973parallel}, and Brent-Kung\citep{brent1982regular} together with generated variants. A categorical diffusion model corrupts and denoises the grid, with a Transformer conditioned on the diffusion step and on a normalized target depth and operator count. Conditioning steers exploration toward a region of the design space, while actual depth and operator count are recomputed from every generated architecture (Appendix~\ref{app:prefix-grid}).

\vspace{2pt}\noindent\textbf{Formal verification and implementation evaluation.}
Every generated architecture is first checked for structural validity, which holds
when the decoded graph is acyclic, combines only compatible intervals, and supplies
the complete carry interval at every output position (Appendix~\ref{app:prefix-grid}). Functional correctness is then verified exhaustively. The architecture is emitted as structural Verilog and compared against $s=a+b$ in a miter construction, and a SAT solver decides whether $\exists\,a,b:s_{\mathrm{candidate}}(a,b)\neq a+b$. The validity predicate $g$ holds only when this query is unsatisfiable, certifying correctness over all $2^{64}$ input pairs. The proof is repeated after technology mapping, confirming that implementation transformations preserve functionality.

Performance is evaluated under a NanGate45 implementation flow that preserves the generated structure through technology mapping and reports delay $D(x)$ and area $A(x)$ after placement-based timing analysis, so the objective is $f_1(x)=(D(x),A(x))$. Novelty uses a canonical schedule-independent fingerprint, so changes in placement or scheduling do not register as architectural novelty, while changes in prefix dependencies do. Appendix~\ref{app:prefix-flow} lists the flow stages.

\vspace{2pt}\noindent\textbf{Search control for structured architectural spaces.}
This space differs from analog topology synthesis because validity is highly
structured and every decoded architecture is efficiently checkable. Under matched
seeds, depth-size targets, and a 192-sample budget, conditional diffusion sampling
beats SMC on hypervolume ($67.8$ vs.\ $53.6$) and best delay ($0.342$ vs.\
$0.362$\,ns), contributing 7 of the 10 merged front points. SMC's resampling
collapses its 192 particles to 7 distinct architectures, so most of its budget
re-evaluates copies. Prefix-adder runs omit mid-generation resampling,
using conditional sampling and $\gamma=0.3$ mutation.

\begin{figure}[t]
\centering
\includegraphics[width=0.9\columnwidth]{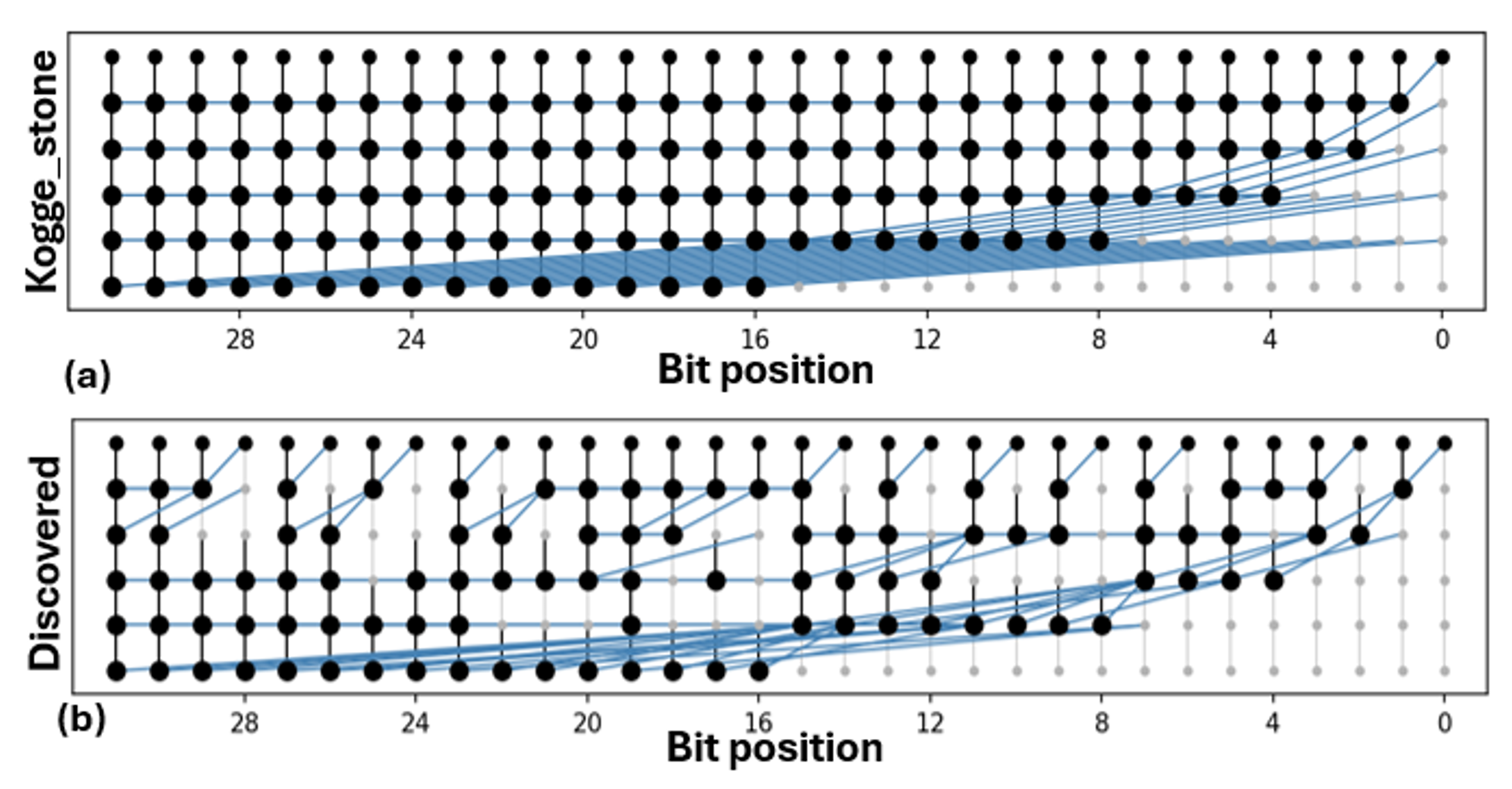}
\caption{The discovered adders are different \emph{wirings} rather than
redrawings of the textbook structures. (a) Kogge-Stone: $0.362$\,ns,
$457\,\mu$m$^2$. (b) Discovered: $0.302$\,ns, $373\,\mu$m$^2$.}
\label{fig:prefix-structures}
\end{figure}

\vspace{2pt}\noindent\textbf{Discovery of improved verified architectures.}
Across 384 zero-shot samples from six depth-size conditions, every sample decodes to a legal prefix graph, and 325 unique architectures are absent from the training corpus. The final set contains 16 novel adders, all formally verified before and after technology mapping. The fastest discovered architecture achieves $0.302$\,ns delay at $373\,\mu\mathrm{m}^{2}$, improving on Kogge-Stone by $17\%$ in delay and $18\%$ in area simultaneously (Table~\ref{tab:prefix}). It removes 33 of the 129 prefix operators used by Kogge-Stone by reducing intermediate dependencies rather than rescheduling the same structure, and Figure~\ref{fig:prefix-structures} shows that the discovered adders are distinct wirings rather than redrawings of textbook designs. Other discovered architectures occupy different regions of the area-delay trade-off.

Absence from the training corpus, formal equivalence over all inputs, and improvement over a mature hand-designed architecture under a physical flow hold simultaneously, which performance optimization alone does not deliver.

\begin{table}[t]
\centering
\small
\setlength{\tabcolsep}{1.7pt}
\renewcommand{\arraystretch}{1.05}

\begin{tabular}{@{}llrrrrrr@{}}
\toprule
\textbf{Method} & \textbf{Design} & \textbf{D} & \textbf{Ops}
& \textbf{FO} & \textbf{Delay} & \textbf{Area} & \textbf{ADP} \\
& & & & & \textbf{(ns)} & \textbf{($\mu$m$^2$)} & \\
\midrule
Textbook
& Ripple       & 31 & 31  & 1  & 1.219 & 175 & 213.3 \\
& Brent--Kung  & 8  & 57  & 5  & 0.464 & 264 & 122.6 \\
& Sklansky     & 5  & 80  & 16 & 0.406 & 322 & 130.9 \\
& Kogge--Stone & 5  & 129 & 5  & 0.362 & 457 & 165.4 \\
\midrule
ArithTreeRL
& Fastest    & 6 & 73 & 16 & 0.342 & 320 & 109.6 \\
& Balanced   & 6 & 69 & 16 & 0.356 & 294 & 104.6 \\
& Min.\ area & 8 & 55 & 6  & 0.477 & 250 & 119.3 \\
\midrule
PrefixGPT
& Min.\ ADP   & 10 & 57  & 6 & 0.516 & 269 & 138.8 \\
& Min.\ delay & 5  & 129 & 5 & 0.362 & 457 & 165.4 \\
\midrule
\textbf{SteerGenSE}
& Fast     & 5 & 96 & 11 & \textbf{0.302} & 373 & 112.8 \\
& Balanced & 7 & 63 & 7  & 0.364 & 280 & \textbf{102.0} \\
& Compact  & 7 & 58 & 6  & 0.406 & 262 & 106.4 \\
\bottomrule
\end{tabular}

\caption{32-bit prefix adders under a common placed-and-timed NanGate45
flow. D, Ops, FO, and ADP denote prefix depth, operator count, maximum
internal fanout, and area--delay product. ArithTreeRL
\citep{lai2024scalable} and PrefixGPT \citep{ding2026prefixgpt} are evaluated
under the same flow. All designs are formally verified and absent
from the corpus.}
\label{tab:prefix}
\end{table}
\subsection{Analog Amplifier Topologies}
\label{sec:analog}

The second instantiation replaces exhaustive verification with simulation-based evaluation. Analog topology discovery is harder to certify and more expensive to evaluate, which is where trajectory-aware steering has its largest effect.

\begin{figure}[t]
\centering
\includegraphics[width=0.75\columnwidth]{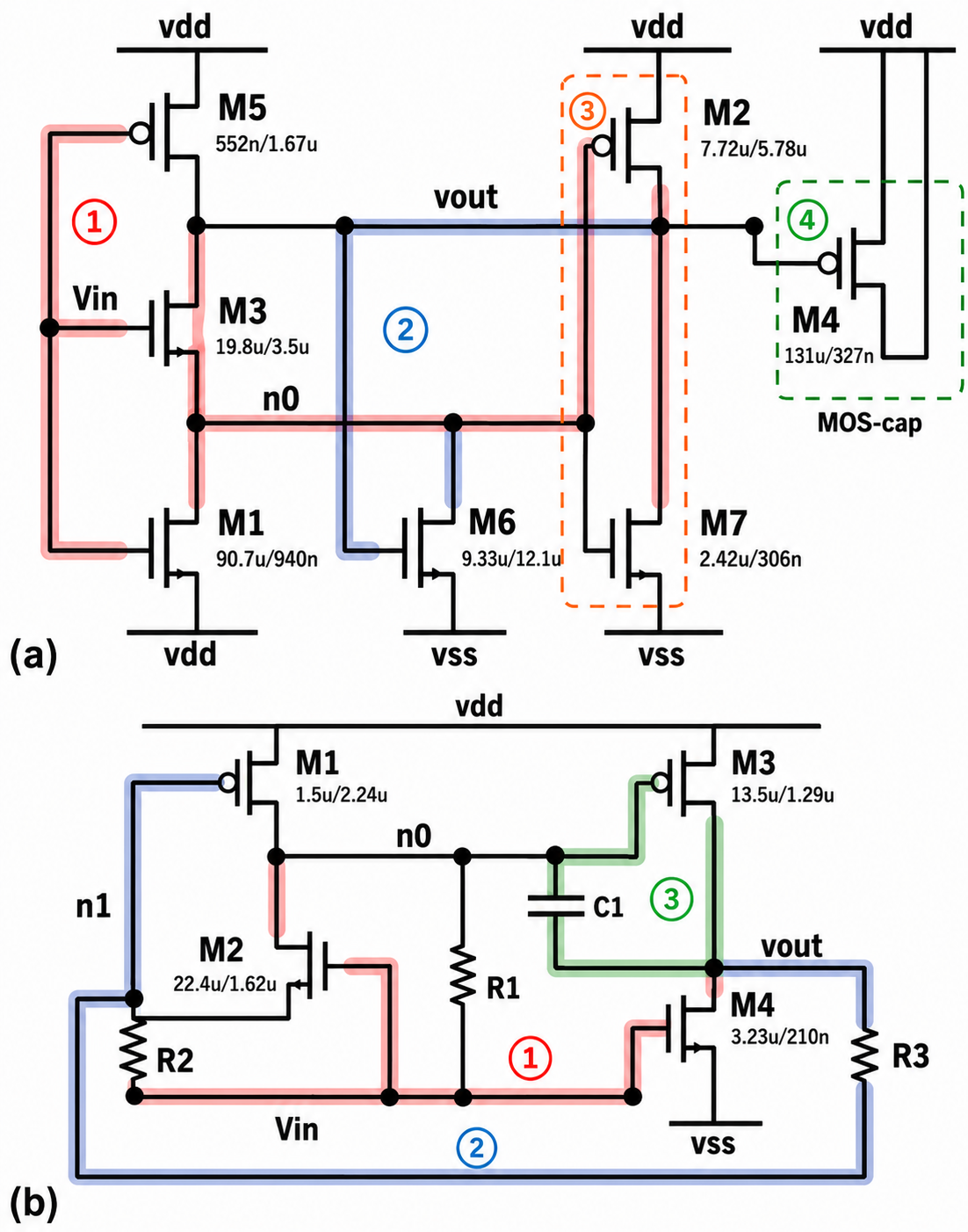}
\caption{Validated amplifiers discovered by diffusion-based search.
In (a), labels mark: \textcircled{1} reinforcing input paths through
M3/M5 and M1-$n_0$; \textcircled{2} M6-mediated regenerative feedback;
\textcircled{3} the M2-M7 inverter gain core; and \textcircled{4} M4
repurposed as a MOS capacitor. In (b), they mark:
\textcircled{1} reinforcing direct and $n_0$-mediated input paths;
\textcircled{2} R2-R3 input--output feedback through $n_1$; and
\textcircled{3} C1 internal compensation.}
\label{fig:discovered-circuits}
\end{figure}

\vspace{2pt}\noindent\textbf{A circuit representation suitable for discovery.}
Discovery here requires deterministic decoding because a candidate that cannot be reconstructed exactly cannot be evaluated. We use a terminal-aware incidence representation that preserves the terminal roles distinguishing electrically different structures. Diffusion operates over a motif-level graph with a vocabulary of 261 recurring subcircuits mined automatically from the training corpus rather than specified by hand. The corpus is derived from the public AnalogGenie corpus \citep{gao2025analoggenie}. Fixed connectivity within each motif guarantees deterministic decoding, eliminates decoder-invented connections, and improves matched raw SPICE convergence by roughly 3-4$\times$ (see Appendix~\ref{app:circuit-rep} for details).

The representation provides decodability, whereas the diffusion prior provides feasibility. Table~\ref{tab:prior_operator_ablation} separates these contributions because every arm uses the same motif vocabulary. Replacing the prior with random initialization over this vocabulary reduces SPICE-scoreable offspring from 28 to 17 and Pareto-front topologies from 5 to 3; replacing $K_{\gamma}$ with random mutation reduces both to zero. Thus, deterministic decoding allows candidates to reach simulation but does not ensure they survive it.

\vspace{2pt}\noindent\textbf{Circuit prior and external evaluation.}
The prior is a discrete diffusion model over the motif-level representation. A graph Transformer predicts the clean-graph distribution through the reverse process, conditioned on circuit family and design requirements. Each generated macro-graph is decoded into a SPICE netlist and evaluated. A structural gate rejects electrically impossible candidates, including missing supply rails or floating terminals (Appendix~\ref{app:circuit-rep}). Because topology alone does not determine performance, surviving candidates undergo inner NSGA-II sizing, with each simulated in \texttt{ngspice}. Novelty is measured using Weisfeiler-Lehman graph similarity\citep{shervashidze2011weisfeiler}.

\vspace{2pt}\noindent\textbf{Search control under circuit discovery geometry.}
Circuit discovery combines highly constrained structural generation with expensive
simulation. Direct gradient-based steering of diffusion trajectories helps little
because intermediate representations do not correspond to valid circuit
transformations. Health-based filtering alone raises structural validity but leaves
Pareto coverage near the unsteered baseline. We therefore use trajectory information
to allocate exploration and evaluation effort, which cuts wasted simulations while
leaving the external evaluator as the final authority. Under a matched budget,
conformal-health SMC raises structural validity from $24.2\%$ to $72.3\%$ and Pareto
hypervolume from $20.5$ to $27.1$, with the full sampler comparison in
Appendix~\ref{app:steering-ablation} (Table~\ref{tab:steering-ablation}).

\begin{figure}[t]
\centering
\includegraphics[width=\columnwidth]{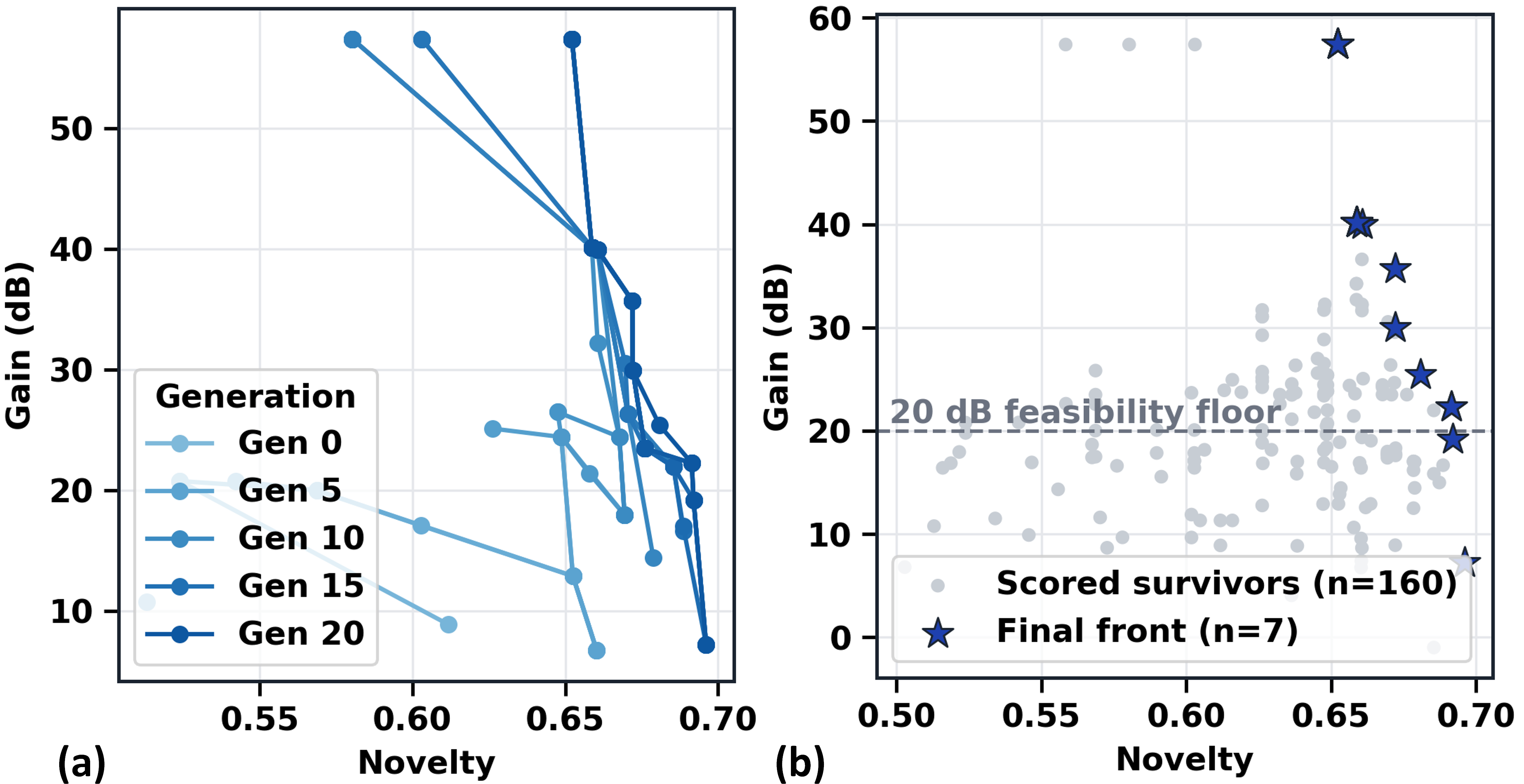}
\caption{Evolution of the novelty-gain Pareto front.
(A) Generation-wise NSGA-II fronts, showing the best gain improving from
10.8 to 57.4\,dB while later generations expand novelty coverage.
(B) Final scored population and non-dominated front; faint points denote
SPICE-scored candidates, and the solid staircase indicates the final Pareto
front. The best design found in this run achieves 57.4\,dB raw gain.}
\label{fig:pareto-evolution}
\end{figure}

\vspace{2pt}\noindent\textbf{Discovery procedure.}
The search maintains a population of $\mu=48$ designs over 20 generations. The
initial population is sampled from the learned prior, screened by the structural
gate, sized by an inner NSGA-II loop, simulated in \texttt{ngspice}, and ranked on
performance and novelty. Each later generation applies $K_\gamma$ to Pareto-ranked
parents, re-noising and denoising them into structurally related offspring. Parents
and offspring are pooled, evaluated through the same pipeline, and Pareto-selected
into the next population. The returned Pareto set spans the performance-novelty
trade-off. Figure~\ref{fig:pareto-evolution} traces one such run. Best raw gain
climbs from $10.8$\,dB in the prior-sampled initial population to $57.4$\,dB by
the final generation, and later generations widen novelty coverage instead of
crowding the highest-gain region, which is the behavior Pareto selection is
meant to produce. Aggregate results below pool three runs, so the best verified
gain there exceeds this single-run maximum.

\vspace{2pt}\noindent\textbf{Discovery of previously unseen circuit organizations.}
Across three runs, the framework proposes 2,432 topologies, of which 1,277 pass
simulation and 22 reach a Pareto front. After removing convergence-dependent
solutions, canonically normalizing device connections, and independently
re-sizing and re-simulating, seven amplifier topologies remain validated
discoveries whose performance persists across sizing solutions. They are absent
from the training corpus, structurally distinct under Weisfeiler-Lehman
comparison, and span gains of $21.9$-$66.1$\,dB, bandwidths of
$72.9$\,kHz-$207$\,MHz, powers of $16.2$-$696$\,mW, and area proxies of
$57.7$-$471$\,$\mu$m$^2$, so the search reaches diverse performance-novelty
regions rather than one template. The discovered topologies contain noncanonical mechanisms not prescribed to the
model, including multiple reinforcing input-to-output paths and a regenerative
loop between an internal node and the output that offers a plausible route to
enhanced small-signal gain. Figure~\ref{fig:discovered-circuits} annotates these
and Appendix~\ref{app:mechanisms} traces them device by device. Diffusion-based
search therefore assembles familiar device-level effects into previously unseen
amplifier organizations. Table~\ref{tab:analog-controlled-comparison} compares
these results with prior analog generators under a common PDK and SPICE flow.

\subsection{Continuous Spatial Organization}
\label{sec:place}

The third instantiation tests whether the discovery formulation transfers from symbolic to continuous domains, rather than serving as an EDA benchmark. Large memory and compute macros are placed on a chip floorplan. Placement introduces no new functionality, so the target is a spatial organization absent from the corpus that improves implementation objectives on unseen designs.

A placement with $n$ macros is $x_0\in\mathbb{R}^{n\times2}$ containing normalized lower-left coordinates, and a continuous diffusion model with $T=200$ steps learns to recover $x_0$ from $x_t$. The denoiser is a Transformer with one token per macro carrying the noisy coordinate, normalized footprint, and log weighted degree. All features are defined relative to the input design, which permits application to unseen netlists. Netlist connectivity is incorporated through attention-based conditioning, detailed in Appendix~\ref{app:attention-bias}.

\begin{table}[t]
\centering
\small
\setlength{\tabcolsep}{1.8pt}
\renewcommand{\arraystretch}{1.05}

\begin{tabular}{@{}lcccccc@{}}
\toprule
\textbf{Design} & \textbf{Macros} & \textbf{Auto} & \textbf{Expert}
& \textbf{ChipDiff.} & \textbf{Ours} & \textbf{Ratio} \\
\midrule
nvdla$^{\dagger}$
& 128 U & 14.78 & 8.34 & 10.03 & 10.01
& $\mathbf{0.68\times}$ \\

ariane136$^{\dagger}$
& 136 U & 4.16 & 3.71 & 3.68 & 3.80
& $\mathbf{0.91\times}$ \\

bp\_fe\_top$^{\S}$
& 11 M & 1.69 & -- & 1.68 & 1.54
& $\mathbf{0.91\times}$ \\

bp\_be\_top$^{\S}$
& 10 M & 2.32 & -- & 2.29 & 2.16
& $\mathbf{0.93\times}$ \\

mempool\_tile$^{\ddagger}$
& 20 M & 3.06 & 2.98 & 3.21 & 3.02
& $0.99\times$ \\

bp\_multi\_top$^{\S}$
& 26 M & 2.88 & -- & 3.20 & 2.91
& $1.01\times$ \\

black\_parrot$^{\S}$
& 24 M & 6.91 & -- & 7.26 & 7.00
& $1.01\times$ \\

swerv\_wrapper
& 28 M & 3.73 & -- & 3.95 & 3.80
& $1.02\times$ \\

bp\_quad
& 220 M & 37.81 & -- & 41.00 & 40.40
& $1.07\times$ \\
\bottomrule
\end{tabular}

\caption{Legalized HPWL in M$\mu$m. Auto denotes the OpenROAD macro
placer; Expert denotes a human placement where available; ChipDiff. denotes
ChipDiffuser \citep{lee2024chip}; and Ours denotes the best SteerGenSE
placement. Ratios are relative to Auto. U and M denote uniform and mixed-size
macro sets.}
\label{tab:place}
\end{table}

\vspace{2pt}\noindent\textbf{Results and limitations.}
Table~\ref{tab:place} evaluates nine designs from the TILOS MacroPlacement and
OpenROAD-flow-scripts suites \citep{cheng2023assessment,ajayi2019toward}; a
zero-shot placement is one denoising pass from the prior, with no search loop. On
\texttt{nvdla} all 16 zero-shot samples beat the OpenROAD Auto placer, and search
reaches $0.68\times$ Auto while improving overflow and timing. Trained without
\texttt{ariane}, the model reaches $0.91\times$ on \texttt{ariane136}, and four
further held-out designs improve on Auto. Quality depends on the representation:
coordinates are normalized to each macro's legal region, so the denoiser must learn
a size-dependent mapping, and uniform-footprint designs reach $0.68$-$0.91\times$
while mixed-size sets degrade to $1.07\times$ on \texttt{bp\_quad}, since the
attention bias captures connectivity but not aspect ratio. Search control also
differs: wirelength and overlap gradients reach $0.66\times$ over 16 zero-shot
samples (Table~\ref{tab:place-steering}), whereas SMC reduces diversity and
conformal pruning on legalization displacement does not help, so search control
must match the geometry of the design space.

\begin{table*}[t]
\centering

{\small
\setlength{\tabcolsep}{7pt}
\renewcommand{\arraystretch}{1.0}

\begin{tabular}{llcccc}
\toprule
Domain & Metric
& \shortstack{Diff prior +\\$K_\gamma$ mutation}
& \shortstack{Diff prior +\\random mutation}
& \shortstack{Random init.\ +\\$K_\gamma$ mutation}
& \shortstack{Random init.\ +\\random mutation} \\
\midrule

\multirow{5}{*}{\shortstack[l]{\emph{Analog}\\\emph{op-amp}\\384 cand.}}
  & Unique topologies
    & 163 & 166 & 161 & 147 \\
  & SPICE-scoreable offspring
    & \textbf{28} & 0 & 17 & 0 \\
  & Final Pareto-front topologies
    & \textbf{5} & 0 & 3 & 0 \\
  & Best raw gain (dB)
    & \textbf{20.8} & -- & 16.5 & -- \\
  & Wall clock (min)
    & 52 & 6 & 21 & 0.8 \\

\midrule

\multirow{4}{*}{\shortstack[l]{\emph{Placement}\\\emph{(nvdla)}\\448 cand.}}
  & $\Delta$HV
    & $\mathbf{+0.0254}$ & $+0.002$ & $+0.018$ & $+0.002$ \\
  & Unique legalized placements
    & 192 & 192 & 156 & 175 \\
  & Best HPWL (M$\mu$m)
    & \textbf{10.008} & 10.064 ($=$ seed) & 10.96 & 13.2 \\
  & Overflow
    & 260 & 295 ($=$ seed) & \textbf{158} & 295 ($=$ seed) \\

\midrule

\multirow{6}{*}{\shortstack[l]{\emph{Prefix}\\\emph{adder}\\288 cand.}}
  & HV
    & \textbf{371.8} & 371.2 & 332.5 & 304.9 \\
  & Unique topologies
    & 239 & \textbf{247} & 203 & 187 \\
  & Front size
    & \textbf{22} & 8 & 8 & 8 \\
  & Best delay (ns)
    & 0.3196 & \textbf{0.3120} & 0.3774 & 0.3889 \\
  & Best area ($\mu$m$^2$)
    & \textbf{262} & 264 & 284 & 313 \\
  & Textbook designs dominated
    & \textbf{2} & \textbf{2} & 0 & 0 \\

\bottomrule
\end{tabular}
}

\caption{Prior/operator ablation at matched budgets, replacing either or both
learned components with random initialization or mutation. Prefix-adder HV
uses the delay--area Pareto front with a fixed reference; placement
$\Delta$HV is relative to the initial population. ``--'' indicates no
scoreable result; overflow is OpenROAD global-routing overflow, where lower
is better.}
\label{tab:prior_operator_ablation}
\end{table*}

\begin{table*}[t]
\centering

{\small
\setlength{\tabcolsep}{3.2pt}
\renewcommand{\arraystretch}{1.0}

\begin{tabular}{lccc|ccccc}
\toprule

Method
& \multicolumn{3}{c|}{Generation quality}
& \multicolumn{5}{c}{Common-PDK evaluation by us} \\

\cmidrule(lr){2-4}
\cmidrule(l){5-9}

&
\shortstack{Valid\\(\%) $\uparrow$}
&
\shortstack{V\&N\\(\%) $\uparrow$}
&
\shortstack{Exact\\(\%) $\downarrow$}
&
\shortstack{Feasible yield\\(\%) $\uparrow$}
&
\shortstack{Best gain\\(dB) $\uparrow$}
&
\shortstack{Best GBW\\$\uparrow$}
&
\shortstack{Power\\(mW) $\downarrow$}
&
\shortstack{SPICE calls\\$\downarrow$}
\\

\midrule

AnalogGenie~\cite{gao2025analoggenie}
& 82.0$^{u}$
& 78.0$^{u}$
& 3.2$^{u}$
& 9/978 (0.9\%)
& 47.1
& 1.3\,MHz
& 11.8
& 417k
\\

AnalogToBi~\cite{kim2026analogtobi}
& \textbf{97.8}$^{p}$
& 89.9$^{p}$
& \textbf{0.0}$^{p}$
& N/A
& N/A
& N/A
& N/A
& N/A
\\

AnalogCoder~\cite{lai2025analogcoder}
& 57.3$^{\mathrm{AG}}$
& 8.9$^{\mathrm{AG}}$
& --
& 4/11 (36.4\%)
& 55.1
& 28.1\,MHz
& \textbf{3.9}
& \textbf{32k}
\\

\midrule

Ours, zero-shot
& 94.3
& 94.3
& 0.0
& 14/512 (2.7\%)
& 56.8
& 184\,MHz
& 78.3
& 66k
\\

\textbf{Ours, full search}
& 96.1
& \textbf{96.1}
& \textbf{0.0}$^{u}$
& \textbf{99/1,008 (9.8\%)}
& \textbf{66.1}
& \textbf{493\,MHz}
& 49.7
& 577k
\\

\bottomrule
\end{tabular}
}

\caption{Comparison with prior generators. V\&N denotes
valid-and-novel designs. Superscripts indicate values reported in the
original paper ($^{p}$), measurements under our PDK and SPICE flow
($^{u}$), and values reported or re-evaluated by AnalogGenie
($^{\mathrm{AG}}$).}
\label{tab:analog-controlled-comparison}
\end{table*}

\section{Ablations and Analysis}
\label{sec:ablations}

\vspace{2pt}
\noindent\textbf{When mid-generation steering helps.}
\label{sec:steering-when}
The two properties in Section~\ref{sec:method-steering} explain the
domain-dependence. Analog evaluation is far costlier than generation, since each
candidate runs an \texttt{ngspice}-intensive sizing loop, and forecast damage
predicts failure, so pruning raises structural validity from $24.2\%$ to
$72.3\%$, cuts simulations per scored candidate from $17.1$ to $4.8$, and lifts
hypervolume from $20.5$ to $27.1$ (Appendix~\ref{app:steering-ablation}).
Prefix-adder evaluation takes milliseconds, so diversity dominates, and
resampling collapses 192 particles to 7 architectures while dropping
hypervolume from $67.8$ to $53.6$. In placement, legalization repairs decoded
violations and weakens damage prediction, so differentiable wirelength and
overlap proxies guide gradients instead, improving best \texttt{nvdla} HPWL from
$0.88\times$ to $0.66\times$ (Appendix~\ref{app:place-steering}). Steering is thus a
cost-control layer selected from pre-search properties, while the prior,
$K_{\gamma}$, and evaluation stay fixed.

\vspace{2pt}
\noindent\textbf{Are both learned components necessary?}
\label{sec:ablation}
Table~\ref{tab:prior_operator_ablation} replaces the prior, the mutation kernel,
or both with random counterparts at matched budgets; removing either degrades
discovery by an amount set by the representation's structure. With the prior
fixed, random mutation destroys the structure simulation requires: analog
SPICE-scoreable offspring fall from 28 to 0 and Pareto topologies from 5 to 0,
and placement returns the seed. Prefix adders, whose grid keeps random edits
decodable, retain hypervolume and slightly better delay ($0.3120$ vs.\
$0.3196$\,ns) but shrink the front from 22 to 8. Replacing the prior removes the
feasible starting region, so random initialization with $K_{\gamma}$ trails the
full method. Table~\ref{tab:sampling_baseline}
(Appendix~\ref{app:sampling-baseline}) isolates evaluator feedback: it expands
the prefix-adder front from 14 to 22, improves \texttt{nvdla} placement from
$0.847\times$ to $0.68\times$ Auto, and raises verified analog topologies from 3
to 7 ($36.2\to66.1$\,dB).

\section{Conclusion}
\label{sec:conclusion}

This work shows that diffusion models can serve as learned search operators for
artifact discovery beyond the training corpus. The learned distribution provides
a feasible search space, diffusion dynamics provide transitions between
artifacts, and external evaluators remain the authority for feasibility and
quality. Across three electronic design spaces, the framework discovers
formally verified prefix adders improving delay and area over Kogge-Stone,
previously unseen amplifier topologies validated through independent
simulation, and competitive macro placements on unseen designs. Limitations
include evaluator cost, representation dependence, and policy-specific
conformal calibration. Overall, generative models can enable algorithmic
discovery by learning not only where valid artifacts exist, but how they evolve.

\bibliography{references}

\end{document}